\documentstyle[11pt,newpasp,twoside]{article}
\def\edcomment#1{\iffalse\marginpar{\raggedright\sl#1\/}\else\relax\fi}
\reversemarginpar

\begin{document}
\title{Study of the Interstellar Medium and Star Formation of a complete sample of 760 isolated galaxies.}
 \author{U. Lisenfeld, L. Verdes-Montenegro, D. Espada, E. Garcia, S. Leon}
\affil{Instituto de Astrof\'\i sica de Andaluc\'\i a (CSIC), Aptdo. 18008 Granada, Spain}

\begin{abstract}
A key problem in astronomy involves the role of the environment in the
formation and evolution of galaxies. In order to answer this question
it is necessary to characterize a reference sample with minimum
influence from the environment, so that its evolution is completeley
determined by nature.  The aim of this project is to provide such
reference, quantifying the ISM properties of a well defined and
statistically significant sample of 760 isolated galaxies.  Based on
optical, H$|alpha$ and infrared luminosities, radiocontinuum emission,
molecular and atomic gas content, compiled from the bibliography or
observed by ourselves, together with POSS-II digitized images, we will
perform a statistical study of the Interstellar Medium (ISM)
 properties as a function of
isolation, and its relation to star formation, morphology and
luminosities, as well as nuclear activity frequency.  This sample will
be different from previous studies by three essential characteristics:
a) strict definition of isolation, b) statistical significance, and c)
complete multiwavelength information concerning the ISM.  It will serve
to evaluate the properties of interacting galaxies, which will be
of special interest to analyse the large amount of data that will be
generated during the next decade for high z galaxies, with the new
instruments to come.  Once the analysis will be finished, the data
will become public by means of a database with free internet acces via
a simple and efficient WEB interface.
\end{abstract}

\section{Introduction}
Although it is widely accepted that galaxy-galaxy interactions stimulate
secular evolutionary effects (e.g. enhanced star formation - SF), morphological
peculiarities including transitions to earlier type, enhanced or diminished 
gas content) the amplitude of the
effects, and processes for accomplishing them, are not well quantified
(Sulentic 1976, 1989; Xu \& Sulentic 1991,Verdes-Montenegro et al. 1998). 
For example there is no clear
consensus  on whether SF enhancement depends on details of an
encounter or the preexisting gas reservoirs available in the galaxies.
It is also difficult to ascribe morphological peculiarities (or the presence
of an AGN) to interaction  because they are also seen in rather isolated
systems. Most of these  uncertainties reflect the lack of a statistically
useful baseline. The goal of this proposal is to create a multiwavelength
database that can serve as a zero point relative to which nurture
effects can be evaluated.

\section{The sample}

We are assembling a multiwavelength database for
a large (n$\sim$760) sample of the most isolated galaxies
in the northern sky. We are using 
the largest reasonably unbiased sample of isolated galaxies, the
Catalog of Isolated Galaxies (CIG: Karachentseva 1973; see also Sulentic
1989) which originally contains n=1052 galaxies, later  reduced to n=893
(Karachentseva 1980). We were able to
derive optical and FIR luminosities for n=760 galaxies that constitute our
full working sample.
A CIG based sample has a number of interesting characteristics and advantages
including:\\
1) ISOLATION - selected on the basis of distance from
nearest neighbors (at least 20$\times$ their diameter). We are in the
process of refining the isolation parameter using the digitized POSS2
atlas.  Local objects will also be excluded due to uncertain isolation
determination.
2) MORPHOLOGY - the CIG is morphologically diverse
to permit statistical studies correlated with Hubble type. Our POSS2
isolation reevaluation is also providing more accurate galaxy types.
3) DEPTH - The CIG samples a deep enough volume of space to allow
us to measure a large part of the radio, IR and optical luminosity
functions.
4) COMPLETENESS - the CIG is reasonably
complete with $<$V/V$_m$$>$=0.42 down to B=15.7. 

\section{A Multiwavelength Database for Our Sample}

Our database will emphasize the properties of the nonstellar material
in the galaxies because this component 
is most sensitive to the effects of external
stimuli. Our database will include 
 observations of neutral (HI), molecular (CO) and ionized
(H$\alpha$) gas, and radiocontinuum emission as an extinction-free tracer of
current SF rate and nuclear activity.
This multiwavelength
survey of the refined CIG sample will allow us to define the mean
statistical properties of the ISM (cold/warm/hot gas and dust) in these
galaxies.
It will allow us to study
the interplay between the SF, ISM and the
environment, the link between central activity (AGN) and
the environment, and the morphology distribution of isolated galaxies.

\end{document}